\documentclass[pra,aps]{revtex4}
\usepackage{latexsym}
\usepackage{bbm}
\usepackage{graphics}
\usepackage{here}
\usepackage{amsmath}
\usepackage{amssymb}
\usepackage{graphicx}
\usepackage{graphics}

\newcommand{\beq}{\begin{equation}}
\newcommand{\eeq}{\end{equation}}
\newcommand{\beqa}{\begin{eqnarray}}
\newcommand{\eeqa}{\end{eqnarray}}

\newcommand{\ket}[1]{\left\vert #1 \right\rangle}

\newcommand{\al}{\alpha}
\newcommand{\ua}{\uparrow}
\newcommand{\da}{\downarrow}
\newcommand{\con}{{\cal C}}

\begin{document}

\title{Entanglement invariant for the double Jaynes-Cummings model}

\author{Isabel Sainz}
\affiliation{School of Information and Communication Technology,
Royal Institute of Technology (KTH), Electrum 229, SE-164 40 Kista,
Sweden.}
\author{Gunnar Bj\"ork}
\affiliation{School of Information and Communication Technology,
Royal Institute of Technology (KTH), Electrum 229, SE-164 40 Kista,
Sweden.}

\date{\today}

\begin{abstract}
We study entanglement dynamics between four qubits interacting
through two isolated Jaynes-Cummings Hamiltonians, via an
entanglement measure based on the wedge product. We compare the
results with similar results obtained using bipartite concurrence
resulting in what is referred to as ``entanglement sudden death''.
We find a natural entanglement invariant under evolution
demonstrating that entanglement spreads out over all of the system's
degrees of freedom that become entangled through the interaction. We
also provide an analysis why certain initial states loose all their
entanglement in a finite time although their excitation and
coherence only vanishes asymptotically with time.
\end{abstract}

\maketitle

\section{Introduction}

Entanglement plays a key role in quantum information processes
\cite{nielsen} and therefore it is important to study entanglement
dynamics in different scenarios. The simplest situation, two-qubit
entanglement dynamics, has been extensively studied in different
contexts
\cite{roa,open,almeida,eberly3,yu,eberly1,eberly2,oliveira,santos,yu2,ficek}.

When interacting with a reservoir, one would naively expect the
entanglement between the two qubits to vanish asymptotically.
However, for certain initial entangled states, the entanglement can
vanish completely in a finite time. This is often referred to as
``entanglement sudden death,'' see for example
\cite{open,eberly3,yu,santos,yu2,eberly1,eberly2} and the references
therein. A recent experimental demonstration was presented in
\cite{almeida} and an open systems analysis was done in
\cite{open,eberly3,yu,santos,yu2}. Nevertheless, if we include the
``reservoirs'' in the studied system and consider the full
entanglement between two non-interacting partitions of the system,
one would expect the entanglement to be preserved, and therefore an
associated entanglement invariant should exist. In addition,
considering the reservoir as part of closed system until the moment
it is ignored (and in mathematical terms ``traced over''), one can
reap insights into both the qualitative and quantitative transfer of
excitation (and associated entanglement) from the atoms to the
reservoir.

An important bipartite interaction is described by the
Jaynes-Cummings (JC) model \cite{JC} that describes, in a concise
and elegant way, the near-resonant interaction between a single
two-level atom and a single-mode quantized field. If the initial
atom-field system contains a single excitation, the system is a
model of a two-qubit system. Since the JC model is excitation number
preserving, the system will always stay within the two qubit Hilbert
space (but of course such a JC model spans only the one-excitation
subspace of the full two-qubit space). However, since this model is
one of the few exactly solvable models in quantum physics it has
been exploited for studying the dynamics of entanglement
\cite{jcentan}.

Recently, a double JC model has been proposed in this context
\cite{eberly1,eberly2}. The model consists of two separate JC-model
systems (atom $A$ interacting only with the cavity field $a$ and
similar for the atom $B$ and the field $b$), where it has been
assumed that the systems are identical. (Note that this model is
applicable to any one-excitation, two-qubit system that is linearly
coupled, e.g., to the experiment in \cite{almeida} where fields
couple pairwise to each other rather than atoms to fields.) A major
reason this particular interaction has been chosen is because it is
local to subsystems $Aa$ and $Bb$. If the atoms, or the fields,
couple to each other, the coupling will alter the entanglement
between the system partitions $Aa$ and $Bb$ in general, and
subsequently, if the fields are traced over, between $A$ and $B$.
The whole point with these studies, however, is to study the
entanglement dynamics between $A$ and $B$ in absence of any
coupling, direct or indirect, between them.

The focus of interest has been the pairwise entanglement dynamics in
terms of the concurrence \cite{concurrence} between the initially
entangled atoms. Through the JC interaction they may become
unentangled through the excitation transfer to the initially
unexcited fields which are traced over after the interaction. In
Ref. \cite{eberly1}, in particular, the authors study entanglement
between the two atoms and they find that for the initial state
$\ket{\psi(0)}=\cos\al\ket{\ua\da}+\sin\al\ket{\da\ua}$, where we
have used the notation $\ket{\ua}_A\otimes\ket{\da}_B=\ket{\ua\da}$,
etc. and $\ket{\ua}$ ($\ket{\da}$) denotes the atom's excited
(ground) state, the concurrence $\con_{AB}$ behave in a harmonic
oscillatory manner. Translating this into a dissipation language,
the entanglement vanishes asymptotically with increasing coupling to
the reservoir. In contrast, the concurrence $\con_{AB}$ of the state
$\ket{\phi(0)}=\cos\al\ket{\ua\ua}+\sin\al\ket{\da\da}$ for certain
values of $\al$, specifically $\vert\tan\al\vert< 1$, falls rapidly
and non-sinusoidally to zero in a finite time and remains zero for
some time. In a dissipation language this means that the
entanglement will vanish in a finite time although both the atomic
excitation and the atomic coherence decay asymptotically.

In order to study the transfer of entanglement between the atom and
the reservoir the authors extend the work in \cite{eberly1} and
study all the 6 concurrences,
$\con_{AB},\con_{Aa},\con_{Bb},\con_{ab},\con_{Ab}$, and $\con_{Ba},
$ for the four-qubit system in \cite{eberly2}. They find that for
the state $\ket{\psi(0)}$ the sum of the concurrences between the
atoms and the fields, $\con_{AB}+\con_{ab}$, is constant under the
JC evolution. This is not so for the state $\ket{\phi(0)}$, but
another function of the six pairwise concurrences and the initial
state (parameterized by $\alpha$),
$\con_{AB}+\con_{ab}+(\con_{Aa}+\con_{Bb})\vert
\tan\al\vert-(\con_{Ab}+\con_{Ba})$, is an entanglement invariant.

In \cite{oliveira} the dynamics for the initial state
$\ket{\psi(0)}$ in an equivalent model, two separately systems
composed by two two-level systems in a dipole-like interaction, is
considered. Here, the authors consider the effect of different
coupling constants in the separately systems. However, the pairwise
entanglement dynamics is expressed in terms of the negativity
\cite{negativiy} and its relation to with the energy transfer.

Motivated by these studies, and in the hope that a more general
entanglement invariant can be found to illuminate the transfer of
entanglement to the system's different parts, we study the same
four-qubit system, where we do not assume that the systems are
identical. To be able to consider all possible bipartite
entanglement we use an entanglement measure introduced by Heydari in
\cite{measure} which is based on the wedge product. We take all the
different possible partitions of the four-qubit system into account,
and we find  an entanglement invariant which does not depend on the
system parameters or on the initial state, provided that it belongs
to a class of pure states denoted ``$X$''-states in \cite{yu},
including both the state $\ket{\psi(0)}$ and $\ket{\phi(0)}$. The
invariant shows that the fields become entangled with all other
parts of the system, all of which is ``destroyed'' (or rather
ignored) when treating the fields as reservoirs.

\section{The model}

Consider a model consisting of two two-level atoms $A$, $B$, each
interacting with a single-mode near-resonant cavity field denoted
$a$ and $b$, respectively. Following \cite{eberly1,eberly2}, we will
assume that each atom-cavity system is isolated and that the
cavities are initially in the unexcited state while the atoms are
initially in an entangled state. The dynamics of this model is given
by the double JC Hamiltonian \beq \hat{H}_{\rm tot}=\hat{H}_A +
\hat{H}_B,\eeq where the Hamiltonians (under the rotating wave
approximation and setting $\hbar = 1$) are \cite{JC},\beq
\label{jch}\hat{H}_k  =  \nu_k (\hat{a}_k^{\dagger}\hat{a}_k+1/2) +
\frac{\omega_k}{2}\hat{\sigma}_z^{k} +
g_k(\hat{a}_k^{\dagger}\hat{\sigma}_-^{k} +
\hat{a}_k\hat{\sigma}_+^{k}),\eeq where $k=A,B$ (where the letter
case is to be interpreted as appropriate), $\nu_k$ is the field
frequency, $\omega_k$ is the transition frequency between the atomic
excited and ground states, and $g_k$ is the coupling constant
between the cavity field and the atom. The field annihilation
operators are $\hat{a}_k$, and $\hat{\sigma}_{\pm}^k$ are the
spin-flip operators defined by
$\hat{\sigma}_-^k\ket{\ua}_k=\ket{\da}_k$,
$\hat{\sigma}_-^k\ket{\da}_k=\hat{\sigma}_+^k\ket{\ua}_k=0$,
$\hat{\sigma}_+^k\ket{\da}_k=\ket{\ua}_k$, and $\hat{\sigma}_z^k$ is
the atomic inversion operator, viz.,
$\hat{\sigma}_z^k\ket{\ua}_k=\ket{\ua}_k$ and
$\hat{\sigma}_z^k\ket{\da}_k=-\ket{\da}_k$. As mentioned in the
Introduction, if we have at most one excitation in each atom-cavity
system, each such system will stay within a two-qubit space. Hence,
since the two atom-cavity systems don't interact, the double JC
model will result in a four qubit state (but again, spanning only a
subspace of the whole four-qubit Hilbert space).

The corresponding evolution operator for the Hamiltonian (\ref{jch})
is \beqa \label{eo}\hat{U}_{k} &= & e^{-it\hat{H}_{k}^0}\left \{
\cos(\hat{\Omega}_kt)\right .  \\
&& \left . -i t \: {\rm sinc} (\hat{\Omega}_kt) \left
[\frac{\Delta_k}{2}\hat{\sigma}_z^{k} +
g_k(\hat{a}_k^{\dagger}\hat{\sigma}_-^{k} +
\hat{a}_k\hat{\sigma}_+^{k})\right ]\right \}, \nonumber \eeqa where
$\hat{H}_{k}^0=\nu_k [ \hat{a}_k^{\dagger}\hat{a}_k+(1
+\hat{\sigma}_z^{k})/2) ]$ is a constant of motion, proportional to
the total number of excitations of system $k$,
$\Delta_k=\omega_k-\nu_k$ is the detuning between the atom and the
cavity for each system, sinc$(x)\equiv x^{-1}\sin(x)$, and
\beq\hat{\Omega}_k=\left \{g_k^2[a_k^{\dagger}a_k+(1
+\hat{\sigma}_z^{k})/2]+\Delta_k ^2/4\right \}^{1/2}.\nonumber\eeq

We will consider that the atoms are initially in an ``$X$''-state,
characterized by a (reduced) atom density operator whose non-zero
elements are found only in the main diagonal and antidiagonal in the
basis $\ket{\ua \ua}, \ket{\ua \da}, \ket{\da \ua}, \ket{\da \da}$.
This class of atom states has the property that the corresponding
two-qubit density matrix preserves the ``$X$'' form when evolving
under the action of certain system dynamics
\cite{yu,yu2,eberly1,eberly2}.  In this case \cite{eberly1,eberly2},
the reason is simple. When an atom transfers its excitation to the
initially empty field, it leaves a signature in terms of the
excitation in the field. Therefore, there cannot exist any coherence
between the states $\ket{\ua \ua}$ and the states $\ket{\ua \da},
\ket{\da \ua}$ unless such coherence existed initially. This is not
the case, by definition, for the ``$X$''-states, and therefore they
will retain their ``$X$'' form under the assumed evolution whose
form was motivated in the Introduction.

As pointed out in \cite{yu}, the ``$X$''-class of states include the
Bell states and the Werner states. Following \cite{eberly1,eberly2}
we will focus on the Bell-like pure states,
 \begin{eqnarray*}
\ket{\phi(0)}=\cos\al\ket{\ua\ua}+\sin\al
e^{i\beta}\ket{\da\da},\\
\ket{\psi(0)}=\cos\al\ket{\ua\da}+\sin\al e^{i\beta}\ket{\da\ua},
\end{eqnarray*}
where $0\leq\al\leq\pi/2$, $0\leq\beta\leq\pi$. (In Sec. \ref{sec:
Invariant} we will consider more general states.) As motivated
above, we will assume that the initial state for the four qubit
model is \beq\ket{\Phi(0)}=\ket{\phi(0)}\otimes\ket{00},\qquad
\ket{\Psi(0)}=\ket{\psi(0)}\otimes\ket{00},\label{instate}\eeq
respectively, where the abbreviated notation
$\ket{0}_a\otimes\ket{0}_b=\ket{00}$ has been used. Notice that Bell
states can be recovered by setting $\al=\pi/4$ and $\beta=0,\pi/2$.
The initial states (\ref{instate}) under the action of the operator
$\hat{U}_A\otimes \hat{U}_B$ evolve as
\begin{eqnarray}
\ket{\Phi(t)}&=&x_1\ket{\ua\ua00}+x_2\ket{\ua\da01}
+x_3\ket{\da\ua10}\nonumber\\\label{phi}&&+x_4\ket{\da\da11}+x_5\ket{\da\da00},\\
\ket{\Psi(t)}&=&y_1\ket{\ua\da00}+y_2\ket{\da\ua00}\nonumber \\
&&+y_3\ket{\da\da10}+y_4\ket{\da\da01},\label{psi}
\end{eqnarray}
where the coefficients for the state (\ref{phi}) are given by
\begin{eqnarray}
x_1 & = & f_A(t)f_B(t) \cos\alpha, \label{eq: x1} \\
x_2 & = & f_A(t)g_B(t) \cos\alpha, \label{eq: x2} \\
x_3 & = & g_A(t)f_B(t) \cos\alpha, \label{eq: x3} \\
x_4 & = & g_A(t)g_B(t) \cos\alpha, \label{eq: x4} \\
x_5 & = & h_A(t)h_B(t)e^{i\beta} \sin\alpha . \label{eq: x5}
\end{eqnarray}
The functions $f_k(t)$, $g_k(t)$, and $h_k(t)$ are given by
\begin{eqnarray}
f_k(t)&=&e^{-i\nu_kt}\left [\cos(\Omega_kt)-i\frac{\Delta_k}{2\Omega_k}\sin(\Omega_kt)\right ],\label{eq: fk}\\
g_k(t)&=&-i\frac{g_k}{\Omega_k}e^{-i\nu_kt}\sin(\Omega_kt),\label{eq: gk}\\
h_k(t)&=&e^{i\Delta_kt/2} \label{eq: hk},
\end{eqnarray}
where the Rabi frequencies are
$\Omega_k=(g_k^2+\Delta_k^2/4)^{1/2}$.

Similarly, the state (\ref{psi}) will have the the coefficients
\begin{eqnarray}
y_1 & = & f_A(t)h_B(t) \cos\alpha, \label{eq: y1} \\
y_2 & = &  h_A(t)f_B(t)e^{i\beta} \sin\alpha, \label{eq: y2} \\
y_3 & = & g_A(t)h_B(t) \cos\alpha, \label{eq: y3} \\
y_4 & = & h_A(t)g_B(t)e^{i\beta} \sin\alpha. \label{eq: y4}
\end{eqnarray}

\section{Entanglement dynamics}
\label{sec: Dynamics} In this section we will analyze the
entanglement evolution in the double JC model. As an entanglement
measure we will use a wedge-product based measure introduced in
\cite{measure}, which, for the two-qubit case, coincides with the
well-known concurrence \cite{concurrence}, and in the multiqubit
case with the entanglement monotones \cite{entmon}. This measure is
defined for any number of subsystems, each having an arbitrary, but
finite, dimension.

By partitioning the total system into two we can compute the
entanglement between these partitions. Consider some partition
composed by $P_1$ with dimension $M$ and $P_2$ with dimension $N$
(note that each partition could contain more than one physical
subsystem). Assume a pure system defined by
\beq\label{purest}\ket{\psi}=\sum_{m=1}^M\sum_{n=1}^N\alpha_{mn}\ket{m}\otimes\ket{n},\eeq
where $\{\ket{m}\}$ and  $\{\ket{n}\}$ are orthonormal bases. In
order to estimate the entanglement between partitions $P_1$ and
$P_2$, we project $\ket{\psi}$ onto the basis states of one of the
partitions. To this end we define the unnormalized state \beq
\ket{\psi_m}=\langle m\vert\psi\rangle.\nonumber\eeq If the system
can be written as a tensor product between a pure state in each
partition, then all states $\ket{\psi_m}$ are parallel. That is,
$\ket{\psi_m}=c_m\ket{\psi_1}$ for all $m=1,\ldots M$, where $c_m$
denotes a $c$-number. If, on the other hand, the pure state
(\ref{purest}) is entangled, then at least two of the vectors, say
$\ket{\psi_m}$ and $\ket{\psi_l}$ are not parallel, and the degree
to which they are not parallel is characterized by the ``area'' the
vectors span. This area is given by the wedge product between the
vectors, but as the wedge product, in general, is signed and
complex, we take the absolute square of the area as a measure of the
nonseparability between these two vectors. The square of the measure
introduced in \cite{measure} can hence be written as the determinant
\begin{displaymath} {\cal A}^2(m,l)=\left\vert\begin{array}{cc}
\langle\psi_m\vert\psi_m\rangle & \langle\psi_m\vert\psi_l\rangle \\
\langle\psi_l\vert\psi_m\rangle & \langle\psi_l\vert\psi_l\rangle
\end{array}
\right\vert.
\end{displaymath}
 Summing all contributions and using symmetry and the fact that the
wedge product between a vector and itself vanish, the entanglement
between $P_1$ and $P_2$ can finally be defined \beq
\label{measure}E_{P_1-P_2}=\frac{1}{2}\sum_{m=1}^M\sum_{l=1}^M {\cal
A}^2(m,l).\eeq
\begin{center}
\begin{figure}
\includegraphics[width=0.45\textwidth]{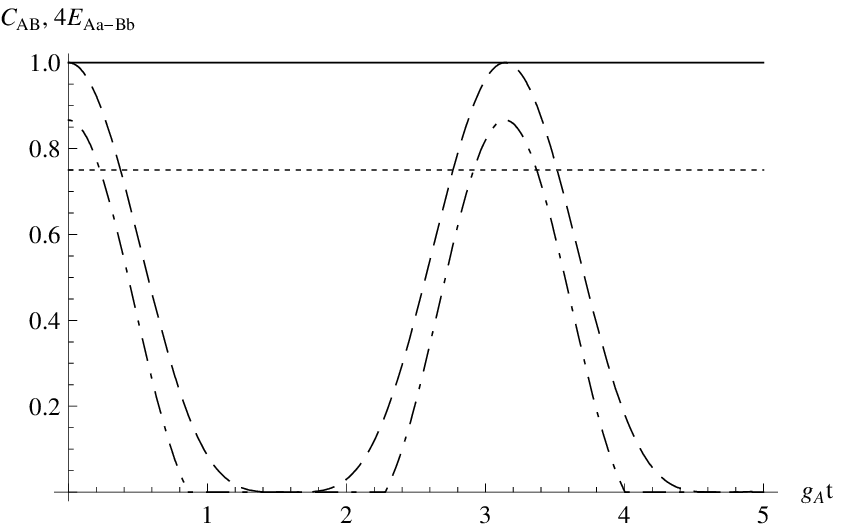}(a)
\includegraphics[width=0.45\textwidth]{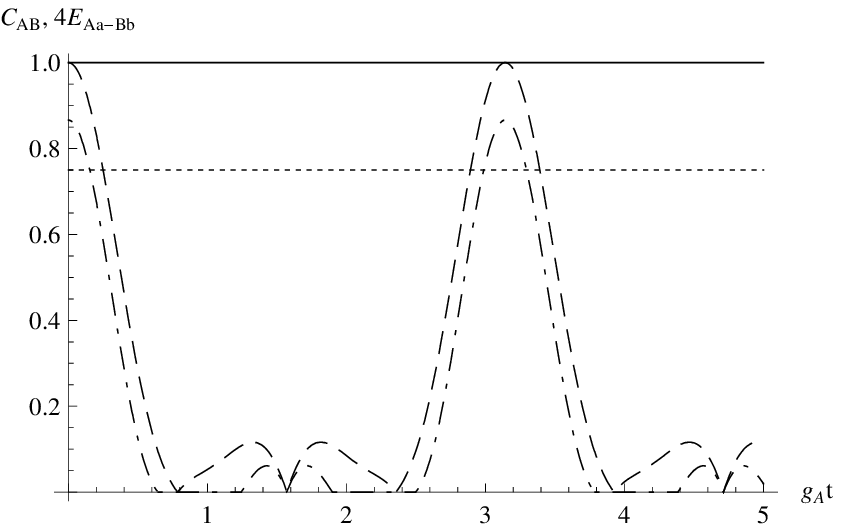}(b)
\caption{(a) The concurrence $\con_{AB}$ between the atoms in state
$\ket{\Psi(0)}$ as a function of the coupling parameter $\Omega_A t$
when $g_A/g_B=1$ for $\al =\pi/4$ (dashed line), $\al =\pi/6$
(dash-dotted line); and $4E_{Aa-Bb}$ for $\al=\pi/4$ (solid line) and $\al=\pi/6$ (dotted line). (b) The concurrence
when $g_B/g_A=2$ for the parameter values $\alpha=\pi/4$ (dashed line) and $\alpha=\pi/6$ (dash-dotted line); and
$4E_{Aa-Bb}$ for $\alpha=\pi/4$ (solid line) and
$\alpha=\pi/6$ (dotted line).}\label{figpsi}
\end{figure}
\end{center}
\begin{center}
\begin{figure}
\includegraphics[width=0.45\textwidth]{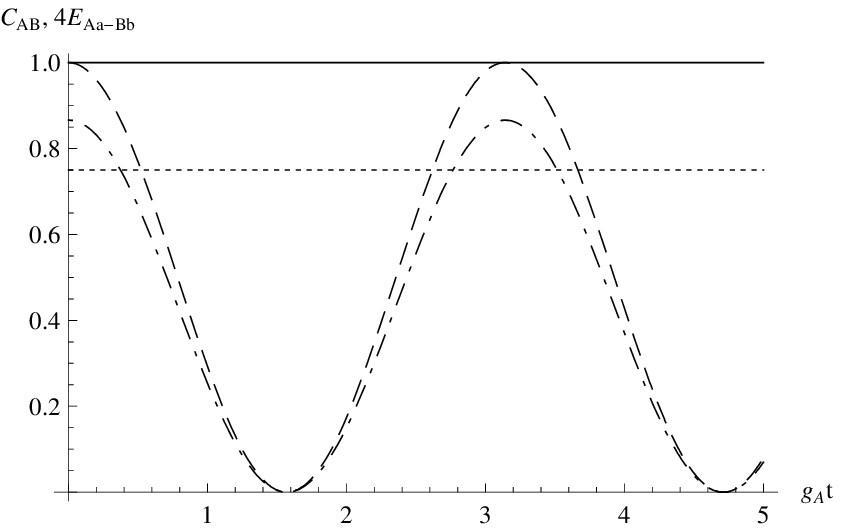}(a)
\includegraphics[width=0.45\textwidth]{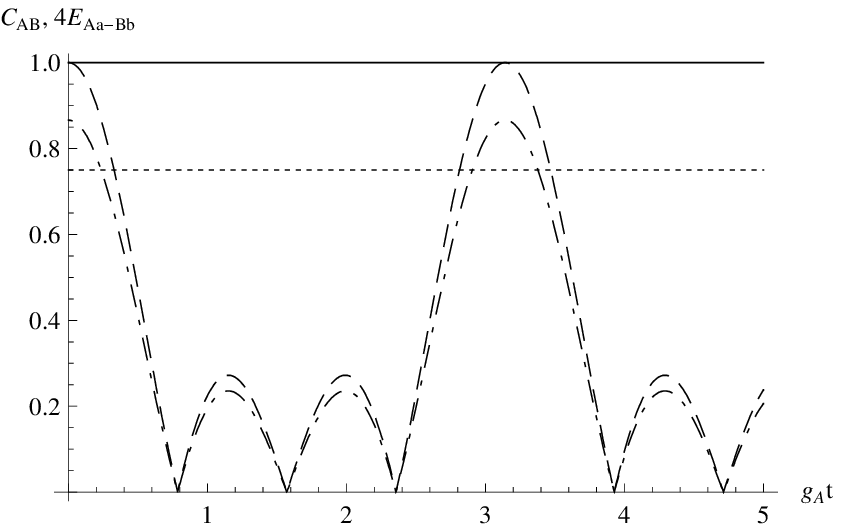}(b)
\caption{ (a) The concurrence $\con_{AB}$ between the atoms in state
$\ket{\Phi(0)}$ as a function of the coupling parameter $\Omega_A t$
when $g_A/g_B=1$ for $\al =\pi/4$ (dashed line), $\al =\pi/6$
(dash-dotted line); and $4E_{Aa-Bb}$ for $\al=\pi/4$ (solid line) and $\al=\pi/6$ (dotted line). (b) The concurrence
when $g_B/g_A=2$ for the parameter values  $\alpha=\pi/4$ (dashed line) and $\alpha=\pi/6$ (dash-dotted line); and
$4E_{Aa-Bb}$ for $\alpha=\pi/4$ (solid line) and
$\alpha=\pi/6$ (dotted line).}\label{figphi}
\end{figure}
\end{center}
In the case of the double-JC model, the possible partitions are: (a)
one qubit - three qubit partitions, $A-Bab$, $B-Aab$, $a-ABb$, and
$b-ABa$; (b) two qubit - two qubit partitions, $Aa-Bb$, $Ab-Ba$, and
$AB-ab$.

In  Figs. \ref{figpsi} and \ref{figphi} we plot the evolution of the
concurrence $\con_{AB}$ between the two atoms $A$ and $B$, and the
evolution of $4E_{Aa-Bb}$ (since $0 \leq \con_{AB} \leq 1$, we use
$4E_{Aa-Bb}$ to scale it to essentially the same range of values as
$C$ may obtain) in the case of exact resonance
($\Delta_A=\Delta_B=0$). Note that while $\con_{AB}$ represent only
the remaining entanglement between the two atoms after the field
states have been traced out, $E_{Aa-Bb}$ represent the entire
bipartite entanglement between the atom-field system $Aa$ and $Bb$.

First we consider different values of $\al$. In Fig. \ref{figpsi}
(a), corresponding to $g_A=g_B$, we can observe that concurrence for
the initial state $\ket{\Psi(0)}$ evolve in a typical oscillating
way between 0 and 1 \cite{eberly1,eberly2,oliveira}, meanwhile
$E_{Aa-Bb}=\sin^2\al\cos^2\al$ is invariant and equals 1/4. In Fig.
\ref{figpsi} (b) we show the time evolution of the concurrence
between the atoms, and that of $4E_{Aa-Bb}$, for $g_B/g_A=2$. When
$g_A\neq g_B$ the concurrence is not evolving in a typical
oscillatory manner as was pointed out in \cite{oliveira}. Meanwhile,
$E_{Aa-Bb}$ depends only on the initial state, namely on $\al$, and
not on the ratio $g_A/g_B$.

Fig. \ref{figphi} shows the evolution of the concurrence and
$4E_{Aa-Bb}$ for the state $\ket{\Phi(0)}$ when both cavity-atom
systems are in exact resonance. In Fig. \ref{figphi} (a) we can
observe the so-called entanglement sudden death for different values
of $\al$. (In the next section we shall discuss this phenomenon in
more detail.) In Fig.\ref{figphi} (b) we show the effect of a different ratio between
$g_A$ and $g_B$.

The difference between the evolution of the two states arises
because state $\ket{\Psi(0)}$ is evolving simultaneously in two
closed manifolds, one consisting of the one-excitation, subsystem
$Aa$ manifold (consisting of states $\ket{\ua \da 0 0}$ and
$\ket{\da \da 1 0}$), the other consisting of the one-excitation,
subsystem $Bb$ manifold (consisting of states $\ket{\da \ua 0 0}$
and $\ket{\da \da 0 1}$). On the other hand, the state
$\ket{\Phi(0)}$ is evolving only in one manifold, consisting of the
states $\ket{\ua \ua 0 0}$, $\ket{\ua \da 0 1}$), $\ket{\da \ua 1
0}$, and $\ket{\da \da 1 1}$. State $\ket{\Phi(0)}$ also has a
ground-state component $\ket{\da \da 0 0}$ but this state does not
evolve.

\section{Dissipative dynamics}
\label{sec: Dissipation}

While the model presented above is closed and does not involve any
dissipation, the atoms' evolution under dissipation can still be
described. As can be seen from Eqs. (\ref{eq: x1})-(\ref{eq: x4}),
(\ref{eq: fk}), (\ref{eq: gk}), and (\ref{eq: y1})-(\ref{eq: y4})
the excitation of the atoms is transferred to the fields in a
monotonic fashion during the time interval where $0 \leq \Omega_k t
\leq \pi/2$. In the resonant case ($\Delta_A = \Delta_B=0
\Rightarrow \Omega_k = g_k$), all the excitation will be transferred
from the atoms to the cavity fields. Formally, during this time
interval, one can then see the fields as a dissipative channel for
the atoms' excitation. On can subsequently map a dissipative
evolution, e.g., spontaneous emission of the atom's excitation
obeying an exponential decay $\propto \exp(- \gamma_k t')$, onto the
JC dynamics by the identification between the times $t$ and $t'$:
$\exp(- \gamma_k t')= \cos^2(\Omega_k t)$. Quite obviously,
$\Omega_k t \rightarrow \pi/2$ correspond to $t' \rightarrow
\infty$. Hence, if the entanglement between the atoms (after the
fields are traced out) become zero in a time $\tau < \pi/(2
\Omega_k)$, then in the dissipative picture it vanishes in a finite
time $t'=\gamma^{-1}\ln(\cos^{-2}[\Omega_k \tau])$. This is indeed
what happens for the state $\ket{\phi(0)}$ as it evolves.

One may then ask why one state's entanglement vanishes in finite
time while the others' does not. The reason is the fundamentally
different way the states decay. The state $\ket{\psi(0)}$ decays
directly into the ground state $\ket{\da \da}$. Whatever excitation
is left in the atoms will still be in a superposition state, and
such a statistical mixture between a Bell state and the ground state
cannot be written as a convex sum of any separable states no matter
to what extent the state has decayed. The state $\ket{\phi(0)}$
decays to the ground state via the intermediate states $\ket{\ua
\da}$ and $\ket{\da \ua}$. As the decay leaves different
``signatures'' in the reservoirs (the states $\ket{0 1}$ and $\ket{1
0}$, respectively), no coherence between these states is
established. When the excitation of these intermediate states is
large compared to the remaining coherence between the states
$\ket{\ua \ua}$ and $\ket{\da \da}$, the state can be written as a
convex combination of separable states so the state is no longer
entangled. This happens when \beq \tan \alpha < \sin^2(\Omega_A t)=
1-\exp(- \gamma_k t') \label{eq: condition} \eeq as pointed out in
\cite{santos,almeida}. If $\cos \alpha < \sin \alpha$, the atomic
excitation is insufficient to excite the intermediate states
$\ket{\ua \da}$ and $\ket{\da \ua}$ to the extent that the
entanglement between the atoms vanishes in a finite time.

If the coupling constants $g_A$ and $g_B$ are different, say that
$g_A > g_B$ as in Fig. \ref{figpsi} (b) and \ref{figphi} (b), the
just presented dissipative picture is only valid as long as $0 \leq
\Omega_A t \leq \pi/2$, where $\Omega_k=g_k$ when the atoms and
cavities are resonant. For times longer than $\pi/(2 g_A)$ the
excitation of atom $A$ starts to revive again in the JC model, a
phenomenon not compatible with dissipation. However, as seen from
the figures, the behavior of the states for times $\Omega_A t <
\pi/2 \Leftrightarrow \Omega_B t < \pi/4 \approx 0.79$ is
qualitatively the same as in the symmetric ($g_A = g_B$) case.

\section{An entanglement invariant}
\label{sec: Invariant}

The form of the chosen interaction Hamiltonian, which does not
include any interaction between subsystems $Aa$ and $Bb$, ensures
that no entanglement is formed between these subsystems that was not
already present in the initial state. Therefore, it is reasonable to
expect an entanglement invariant to exist that measures the net
entanglement between these subsystems. As Fig.\ref{figpsi} and Fig.\ref{figphi} suggest, we can find
the invariant \beq {\cal E}  =
E_{Aa-Bb} \label{inv}\eeq
valid for all parameter values, that is, even for nonresonant
coupling and different atom-cavity coupling ratios.

The introduced measure (\ref{inv}) does not depend on time for the
initial pure ``$X$''-states (\ref{phi}) and (\ref{psi}). The value
of ${\cal E}$ in both cases is \beq {\cal
E}=\sin^2\alpha\cos^2\alpha. \eeq This value is proportional to the
square of the two atoms' initial concurrence $C$ \cite{concurrence}:
\beq 4 {\cal E}= C^2.\nonumber\eeq The result is expected, because
as the Hamiltonian is chosen not to change the entanglement between
$Aa-Bb$, it must remain equal to its initial value at all times.
What is more significant is that for the initial generic state \beqa \ket{\xi_g}&=&c_1\ket{\ua\ua
00}+c_2\ket{\da\ua10}+c_3\ket{\ua\da01}+c_4\ket{\da\da11}+c_5\ket{\da\da00}\nonumber\\
&&+d_1\ket{\ua\da00}+d_2\ket{\da\ua00}+d_3\ket{\da\da10}+d_4\ket{\da\da01},\label{genstate}\eeqa
we find that (\ref{inv}) is still invariant during evolution under the action of
$\hat{U}_A\otimes \hat{U}_B$. Explicitly,
\begin{eqnarray}\label{geninv}
{\cal E}(\ket{\xi_g}) & = &
|c_1c_4-c_2c_3|^2+|c_1d_3-c_2d_1|^2+|c_3d_3-c_4d_1|^2\nonumber\\&&+|c_1d_4-c_3d_2|^2+|c_2d_4-c_4d_2|^2+|c_1c_5-d_1d_2|^2\nonumber\\&&+|c_2c_5-d_2d_3|^2+|c_3c_5-d_1d_4|^2+|c_4c_5-d_3d_4|^2.\end{eqnarray}
It is worth noticing that each cavity-atom in the state $\ket{\xi_g}$ will evolve in the subspace $\{\ket{\ua0},\ket{\da1},\ket{\da0}\}$ under the double JC Hamiltonian. Hence, the state (\ref{genstate}) can be seen as a two-qutrit, pure state. In this case equation (\ref{geninv})is just one ninth of the square of the two-qutrit concurrence introduced in \cite{cereceda}.

It should also be noted that in
the dissipative picture, the state $\ket{\xi_g}$ models coupling to excited reservoirs, and in general
the states cannot, at any time, be written as a product state
between the atoms and the fields. Hence, some entanglement between
atom and fields is already there at the start of the evolution. The
states $\ket{\Phi (0)}$ and $\ket{\Psi (0)}$ are special cases of
$\ket{\xi_g}$ where the atoms couple to
initially empty reservoirs, a relevant but special case.

When we study other partitions in (\ref{measure}), we have seen that for both
states, any single term is zero only at discrete times. This means
that at all times, except a discrete set of times of zero measure,
all parts of the system become entangled in some degree through
excitation transfer. This phenomenon is generic for entangled
systems and has, e.g., been used to entangle subsystems that have
never interacted through so-called entanglement swapping \cite{Pan}.
In a dissipative system, this entanglement spread is of course
detrimental and may lead to complete elimination of entanglement.

\section{Discussion}

In this work we have discussed the entanglement dynamics for two
excited atoms coupled to cavity field-modes through a
Jaynes-Cummings Hamiltonian. We have discussed both the closed
system dynamics and the dynamics if the cavities are viewed as
reservoirs. We have discussed why different initial atomic states
face different fates (asymptotic vs. sudden decay) with respect of
their entanglement under dissipation. We have also shown that as
expected, there exist an entanglement invariant valid for a large
class of states, much larger than the class of so-called ``$X$''-states in the closed system. When treating the fields as
reservoirs (i.e., when tracing over the fields), some of the
entanglement transferred to the cavities is ignored. The state
$\ket{\phi(0)}$ transfers its excitation over a larger set of
distinguishable field states ($\ket{01}$, $\ket{10}$, and
$\ket{11}$) than the state $\ket{\psi(0)}$ (that excites only the
field states $\ket{01}$ and $\ket{10}$), and therefore is is not so
surprising that the former state may loose all of its entanglement
through even a finite dissipation.

Since the proposed (bipartite) invariant is a measure of the entanglement between the two systems $Aa$ and $Bb$,
and the measure is invariant to local unitary transformations, it will be a constant for any pure state, not only for the generic state (\ref{genstate}), but also for states with higher excitation.

\section*{Acknowledgements}
This work was supported by the Swedish Foundation for International
Cooperation in Research and Higher Education (STINT), the Swedish
Research Council (VR), and the Swedish Foundation for Strategic
Research (SSF).

\end{document}